\documentclass[fleqn,usenatbib]{mnras}

\usepackage[normalem]{ulem}

\usepackage{newtxtext,newtxmath}
\usepackage[T1]{fontenc}

\usepackage{graphicx}	% Including figure files
\usepackage{amsmath}	% Advanced maths commands
\usepackage{soul}       % For editing with strikethrough 

\newcommand{\dkl}{\mathcal{D}_{KL}}
\newcommand{\djs}{\mathcal{D}_{JS}}
\newcommand{\djsdens}{\delta_{{JS},\nu}}

\title[Information Content of Biosignatures]{An Information Theory Approach to Identifying Signs of Life on Transiting Planets}

\author[S. Vannah et al.]{
Sara Vannah,$^{1,2}$
Marcelo Gleiser,$^{1}$\thanks{E-mail: mgleiser@dartmouth.edu}
and Lisa Kaltenegger$^{3,4}$
\\
$^{1}$Department of Physics and Astronomy, Dartmouth College, Hanover, NH 03755, USA\\
$^{2}$Atmospheric and Environmental Research, Inc., Lexington, Massachusetts 02421, USA\\
$^{3}$Carl Sagan Institute, Space Science Building 311, Ithaca, NY 14850, USA\\
$^{4}$Cornell University, Astronomy and Space Sciences Building, Ithaca, NY 14850, USA
}

\date{Accepted 2023 October 12. Received 2023 September 21; in original form 2023 July 13}

\pubyear{2023}

\begin{document}
\label{firstpage}
\pagerange{\pageref{firstpage}--\pageref{lastpage}}
\maketitle

\begin{abstract}
Can information theory provide insights into whether exoplanets are habitable? Here we apply information theory to a range of simulated exoplanet transmission spectra as a diagnostic tool to search for potential signatures of life on Earth-analog planets. We test the algorithms on three epochs of evolution for Earth-like planets orbiting a range of host stars. The James Webb Space Telescope and upcoming ground- and space-based missions promise to achieve sufficient high-resolution data that information theory can be applied to assess habitability. This approach provides a framework and a tool for observers to assess whether an exoplanet shows signs of habitability. 
\end{abstract}


%%%%%%%%%%%%%%%%%%%%%%%%%%%%%%%%%%%%%%%%%%%%%%%%%%

%%%%%%%%%%%%%%%%% BODY OF PAPER %%%%%%%%%%%%%%%%%%

\section{Introduction}\label{sec:intro}
%What field knows
Since the observation of the first planet outside our solar system over 5500 exoplanets have been confirmed\footnote{The current number of observed exoplanets can be found on \href{https://exoplanetarchive.ipac.caltech.edu}{NASA's Exoplanet Archive}.}, showing a remarkable diversity of exoplanets orbiting different host stars. Current and future ground- and space-based telescopes promise to vastly expand both the quality and quantity of exoplanet observations, accelerating our ability to search for life in the cosmos. The James Webb Space Telescope (JWST) \citep{gardner2006james} allows for detailed spectroscopy of exoplanets, including rocky, potentially Earth-like worlds. Recent data from JWST was used to identify carbon dioxide, water, sulfur dioxide, and sulfur monoxide in the atmosphere of gas giant WASP-39 b \citep{aher2022, rustamkulov2023early, ahrer2023early, alderson2023early, feinstein2023early}, water in gas giant WASP-96 b \citep{pontoppidan2022jwst}, and carbon dioxide and methane in K2-18 b \citep{madhusudhan2023carbonbearing}, confirming JWST's ability to resolve specific molecular signatures. Soon upcoming missions such as the Atmospheric Remote-sensing Infrared Exoplanet Large-survey (ARIEL) \citep{tinetti2018chemical}, Earth-2.0 \citep{ye2022china} and the ESO Extremely Large Telescope (ELT) \citep{ramsay2020eso} promise to provide more spectroscopic data focused specifically on the search for life. 

However, the challenges of interpreting signs of life and habitability in alien environments remain. Characterization of habitability alone features a slew of complex questions requiring rough inference to categorize a planet as having the potential to host life (see e.g. \cite{seager2013exoplanet, seager2014future, Schwieterman2018, kaltenegger2017characterize}). Without the ability to spatially resolve the surface of exoplanets, spectroscopy of their atmospheres provides a promising avenue for detecting potential biological activity on other planets. We focus on biosignatures first proposed by \citet{Lederberg1965} and \citet{Lovelock1965} as pairs of chemicals in a planetary atmosphere that are out of thermodynamic equilibrium with concentrations unlikely to be sustained by inorganic processes. Rather, the concentrations are sustained by biotic activity on the planet. In particular, these authors suggested the pairs CH$_4$ + O$_2$ and CH$_4$ + O$_3$ as biomarkers. 

%Remaining gap
Increased access to data accelerates the need for quantitative methods to assess the likelihood that an exoplanet hosts life. The ideal method will require as few assumptions as possible about the form that observational signs of life take, minimizing chances of encountering life but being unable to identify it. Proposed methods for these so-called ``agnostic biosignatures'' often utilize complexity measures including through in situ measurements \citep{Guttenberg2021, Marshall2021, Chou2021}, detailed chemical networks in the planet \citep{Wong2023}, and temporal variation in the planetary reflectance spectrum \citep{Bartlett2022}. In this work, we propose an information-theory-based method that quantifies the Earth-likeness of a planetary spectrum without knowing \textit{a priori} what features to look for.

Here we assess the performance of information theory to explore how ``Earth-analog'' an exoplanet is, defined here as having  spectral features with similar syntactic (as opposed to semantic, or meaningful) information content to modern Earth's. This is based on the distance in information-entropic space (known as Jensen-Shannon Divergence, $\mathcal{D}_{JS}$) between the exoplanet's transit spectrum and modern Earth's spectrum (see \citet{vannah2022informational}). We consider planets with Earth-like parameters (radius, mass, surface pressure, temperature, varying atmospheric composition based on Earth's evolution models) observed in different contexts (different host stars and different ages) to explore whether we can identify the biosignature pairs CH$_4$ + O$_2$ and CH$_4$ + O$_3$ through atmospheric spectral signatures. We introduce a quantitative assessment of how ``Earth-analog'' a specific spectral feature is, known as $\djs$ density or simply $\djsdens$. Quantifying the difference in information content between an exoplanet and Earth as a function of wavelength allows us to determine how similar an exoplanet appears to Earth at that wavelength. Capturing the difference in information content of the spectra allows us to assess aspects of the spectral features (such as their shape) that we may not know \textit{a priori} to look for. Clues to differentiating molecules with degenerate absorption bands may be hidden in the shape of spectral features. 

One complicating factor in the search for potentially habitable planets is the variety of contexts an inhabited planet may be observed in. For example, a planet with the potential to host life may be observed too early in its evolution, causing it to look dissimilar to modern Earth. Similarly, an Earth-like planet could be observed orbiting a host star with a  spectral class different from the Sun. This change in spectral irradiance selects which molecules are most rapidly photodissociated in the planet's atmosphere, thereby affecting its chemical composition. Here, we apply our information theory method to a subset of simulations of Earth-like planet transmission spectra at three different stages of their biological evolution (at modern (0Gya), 0.8Gya, and 2.0Gya) and/or orbiting 8 different host stars (F7V, G2V (solar), G8V, K2V, K7V, M1Va, M3Va, and M8Va) from \citet{kaltenegger2020finding} to assess the influence of the context on our results. 

In Section \ref{sec:Data} we describe the dataset; Section \ref{sec:Info} discusses $\djsdens$ and the information method. In Section \ref{sec:Results}, we present key results and demonstrate how the information content of an Earth-like model changes relative to our modern Earth with age and host star. Section \ref{sec:Discussion} discusses limitations and further applications of our analysis. In Section \ref{sec:Conclusions} we summarize our results and conclude that information theory provides a tool to identify absorption features of methane, ozone, molecular oxygen, and water in a transmission spectrum, tracing the impact of life on Earth's atmosphere over time.

\section{Data} \label{sec:Data}
In order to gather as much information as possible from exoplanetary spectra while requiring minimal input knowledge about their spectral appearance, we define an information-entropic spectral method (defined in Section \ref{sec:Info}) to quantify how ``Earth-analog'' a particular transmission spectrum is. We apply this method to simulated data from \citet{kaltenegger2020finding}, which use the well-known \texttt{EXO-Prime} code \citep{kaltenegger2009detecting, rugheimer2018spectra}, a 1D iterative climate-photochemistry atmospheric model coupled with a line-by-line radiative transfer code developed for Earth \citep{traub2002}, adapted to rocky exoplanets \citep{kaltenegger2007} to generate transmission spectra for Earth-like planets. EXO-Prime models high-resolution transmission spectra from the visible to the infrared (0.4---20$\mu$m) at steps of 0.01cm$^{-1}$. The evolutionary stage of an Earth-like planet, as well as the type of host star, impacts the chemical composition of the planetary atmosphere by determining, for example, which molecular species experience photodissociation. To maximize our chances of identifying life on an inhabited planet, it is critical to identify Earth-like planets at different ages and orbiting different classes of host stars. 

We smooth the data to $\lambda/\Delta\lambda=300$ for clarity, using the telescope noise modeling package \texttt{coronagraph} \citep{Lustig2019coronagraph, Robinson2016}, which performs a top-hat convolution to re-bin the spectrum to a lower resolution. The spectra are shown for reference in the Supplemental Material. At this resolution, several features overlap but can be discerned at high-resolution, as detailed in \citet{kaltenegger2020finding}. We consider model spectra for a series of Earth-like planets orbiting F, G, K, and M-type host stars for three epochs in the evolution of life on Earth. We use these simulations to determine a range of expected information contents to categorize an exoplanet as ``Earth-analog.'' The simulations assume an Earth-like planet with Earth mass, radius, pressure, temperature, and model atmospheric composition through geological evolution \citep{kaltenegger2020finding}. The major chemical species included in the simulation are O$_2$, O$_3$, OH, O, H$_2$O, HO$_2$, H$_2$O$_2$, CO$_2$, CO, H$_2$CO, CH$_4$, CH$_3$O$_2$, CH$_3$OOH, CH$_3$Cl, HCl, HOCl, Cl$_2$O$_2$, ClO, ClONO$_2$, SO$_2$, H$_2$S, H$_2$SO$_4$, HSO, HS, H$_2$, H, N$_2$O, NO$_2$, NO$_3$, NO, HNO$_2$, and HNO$_3$. 

\section{Information-Entropic Spectral Method}\label{sec:Info}
\citet{vannah2022informational} explored how information theory can be used to identify Earth-analog planets. We briefly review the method here, explaining how it can be expanded to investigate the information content of particular wavelengths linked to specific 
potential biosignatures.

We begin by defining the modal fraction, $p_\nu$, of the spectrum. This discrete probability density developed first in \citet{Gleiser2012} quantifies the relative weight of a particular wavenumber $\nu$ in the measured spectrum (related to the wavelength, $\lambda$, by $\nu = 1/\lambda$). Since transmission spectra represent the strength of absorption as a function of wavenumber, the modal fraction is the normalized relative weight of a given mode,
\begin{equation}
    p_\nu = \frac{h_\nu}{\sum_\nu h_\nu},
\end{equation}
where $h_\nu$ is the effective height of the atmosphere at a particular wavenumber, $\nu$. The effective height of an exoplanet atmosphere is the difference between the (wavenumber-dependent) transit depth and the geometric transit depth (the transit depth if the exoplanet did not have an atmosphere). It represents the maximum height of the atmosphere that light can penetrate at that wave number. The information content of a spectrum is defined using the Shannon entropy relation \citep{Shannon1948}, 
\begin{equation}
    \hat{H} = {\sum_\nu p_\nu log(p_\nu )}.
\end{equation}
The difference in estimated Shannon entropy, $\hat{H}$, between two modal fractions representing two spectra, $p_\nu$ and $q_\nu$, is closely linked to the Kullback-Leibler Divergence \citep{kullback_information_1951}, $\dkl$, defined as 

\begin{equation}
    \mathcal{D}_{KL}(p||q) = \sum_\nu p_\nu log (\frac{p_\nu}{q_\nu}).
\end{equation}
Since $\dkl$ is not symmetric on exchange of $p_\nu$ and $q_\nu$, it cannot represent a true metric distance in information space  between the two spectra. For example, the difference between the information content of the spectra of Earth and Mars is the same as between Mars and Earth, but the $\dkl$ between them is not. To address this issue, we use the symmetrized 
Jensen-Shannon Divergence \citep{lin1991divergence},  $\mathcal{D}_{JS}$, given by

\begin{equation}
    \mathcal{D}_{JS}(p||q) = \frac{1}{2}\sum_\nu p_\nu log (\frac{p_\nu}{r_\nu}) + \frac{1}{2}\sum_\nu q_\nu log (\frac{q_\nu}{r_\nu}),
\end{equation}
\noindent
where $r_\nu=\frac{1}{2}(p_\nu+q_\nu)$ is the mean of the modal fractions. Since $\djs$ is obtained by summing over all wavenumbers within a certain range, it gives a global numerical distance in information space between the two spectra being compared. As shown in \citet{vannah2022informational},
$\djs$ provides a quantitative measure with minimal input knowledge for evaluating how similar two planets are even without knowing \textit{a priori} what similarities to look for. In log base 2, $\djs$ has units of bits. For a signal that can be expressed in a series of true-false questions---such as a message encoded in logic gates in a computer---the number of bits is the number of true-false questions that need to be answered to convey the message. Each answer (true or false) would have a probability of $\frac{1}{2}$. For modal fractions, each data point in the spectrum has a different weight---or probability---which is less than $\frac{1}{2}$. As a result, the $\djs$ of two modal fractions is less than one bit. In this work, we use natural logarithm rather than log base 2, which gives $\djs$ in units of nats. 

To extract the information divergence between two spectra for a particular biosignature feature, we use the $\djs$-density, $\djsdens$, given simply by the summand of $\djs$,
\begin{equation}
    \djsdens (p_\nu||q_\nu)= \frac{1}{2} p_\nu log (\frac{p_\nu}{r_\nu}) + \frac{1}{2}q_\nu log (\frac{q_\nu}{r_\nu}).
\end{equation}
\noindent
Whereas $\djs$ gives a global measure comparing two spectra, $\djsdens$ provides a local measure, allowing us to search for and directly compare specific spectral signatures, such as those of a specific compound or combinations thereof related to potential biotic activity. Using the information measure as opposed to simply comparing ratios of specific wavenumbers from different spectra offers several advantages. By construction, $\djsdens$ is calibrated to compare different wavenumbers within a joint normalized metric space for the two spectra of interest (obtained through the quantity $r_{\nu}$ that combines the two spectra). This metric space depends explicitly on the whole measured spectra and not on two arbitrary readings for specific compounds obtained from two different spectra.
Due to detector resolution and data accuracy, in practice $\djsdens$ will quantify the spectral divergence at a coarse-grained value averaged over a range of wavenumbers. Using known approximate locations of specific potential biosignatures, $\djsdens$ quantifies how similar the two spectra are at this particular location without assuming prior knowledge about the planetary source. This is critical, as information in the line shape may reveal an enhanced signal of inhabitance.

In what follows, we will use both $\djs$ and $\djsdens$: $\djs$ is sensitive to the full spectrum and can be used to identify how Earth-analog a rocky planet appears, $\djsdens$ can then be used---assuming certain specific biosignatures---to identify \textit{Earth-analog} planets from amongst the pool of potentially habitable ones, as well as to explain the behavior seen in the $\djs$.

\section{Results} \label{sec:Results}
\subsection{How Earth-analog is an exoplanet spectrum?}\label{subsec:res1}
%Our question
In order to apply information theory to determine how Earth-analog a rocky planet appears and to identify potential signatures of life, we first compute the $\djs$ for our simulated Earth-like planets, obtaining  $\djs$  and $\djsdens$ in different contexts. We split the eight star-planet systems into two groups: a hotter group of host stars (F7V, G8V, K2V stars, and the Sun), as well as a cooler group (K7V, M1V, M3V, and M8V stars). Because of the strong effect of the host star on a planet's spectrum, we compare planets orbiting the hotter star systems to modern Earth around our Sun, while planets orbiting the cooler star systems are compared to a modern Earth model around a M1V star. 

\begin{figure}
    \centering
    \includegraphics[width=\columnwidth]{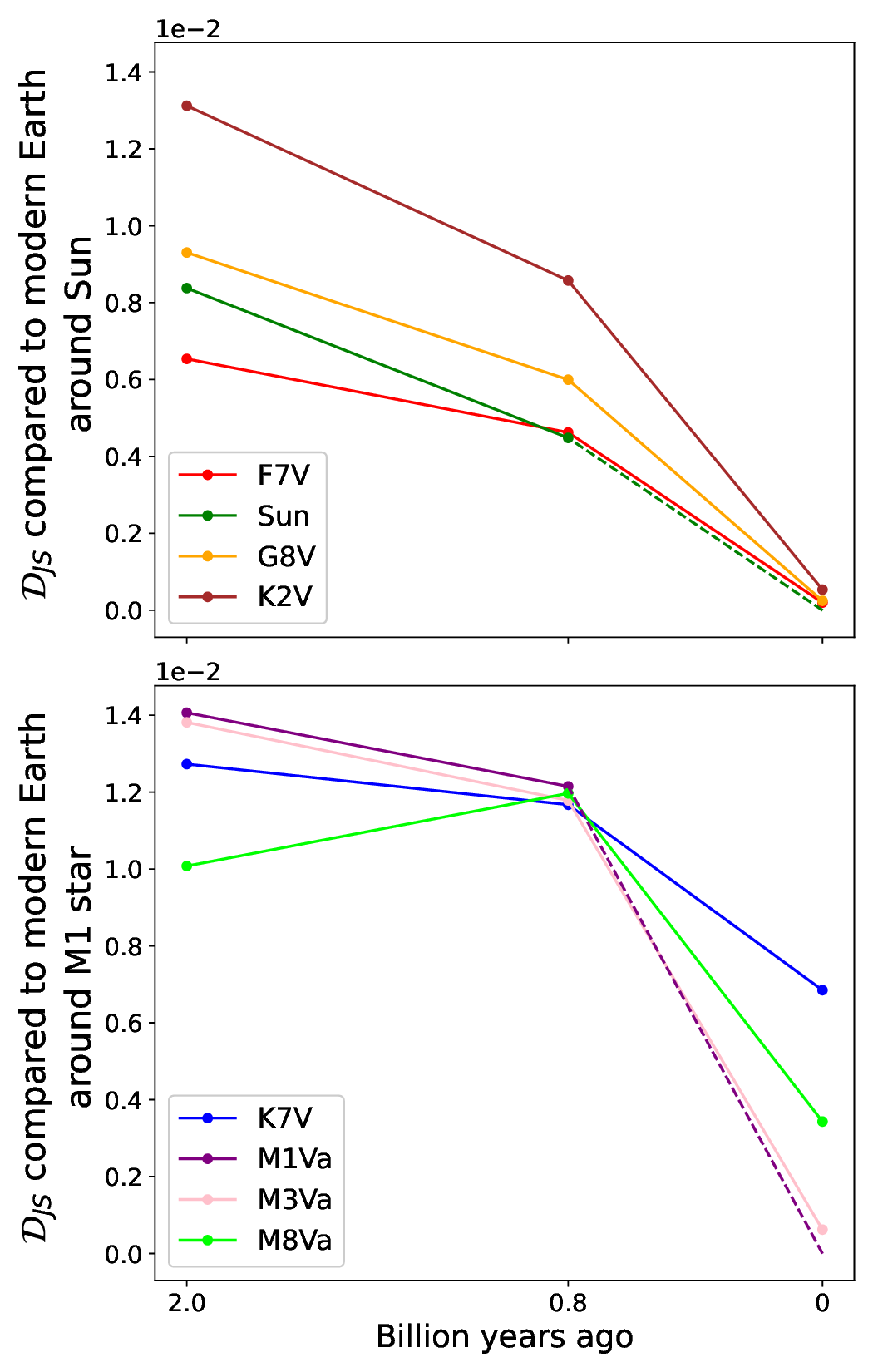}
    \caption{$\djs$ of Earth-like planets through their evolution, orbiting different spectral class host stars (specified in Figure legend). The planets orbiting hotter stars (top) are compared to a modern Earth around the Sun, while the planets orbiting cooler stars (bottom) are compared to a modern Earth model around an M1V star. A small $\djs$ indicates low information loss relative to their comparison planet---meaning that the two planets are most similar. }
    \label{fig:djs_time}
\end{figure}

Figure \ref{fig:djs_time} shows the results for the total $\djs$ for both hotter stars (top) and colder stars (bottom). A lower $\djs$ indicates that the exoplanet's spectrum is more similar to a modern Earth-like planet around its comparison star. For the top plot, this means a lower $\djs$ indicates a planet more similar to a modern Earth around our Sun, for the bottom plot, a planet more similar to a modern Earth around an M1V star. 

The general behavior of the planets follows the expected trends. We find that planets are most similar to their respective Earth-standard when they share Earth's age. This is shown in the far right of each plot, where the $\djs$ for most of the planets is the lowest. The spectra of Earth-like planets orbiting stars closer in temperature to the Sun also result in lower $\djs$ than stars much hotter or cooler than the Sun. 

Note that $\djs$ values for planets of different ages orbiting different host stars can overlap. This illustrates that the modeled planet epoch and the spectral class of the host star do not impact $\djs$ independently. Rather, both must be considered when analyzing the spectrum of an exoplanet. This illustrates the need to compare exoplanet spectra to an Earth-like planet in the appropriate context. Our analysis shows that to identify Earth-analog planets, one needs to use a comparison to a host star that is similar when using information theory. 

\begin{figure*}
    \centering
    \includegraphics[width=1.0\textwidth]{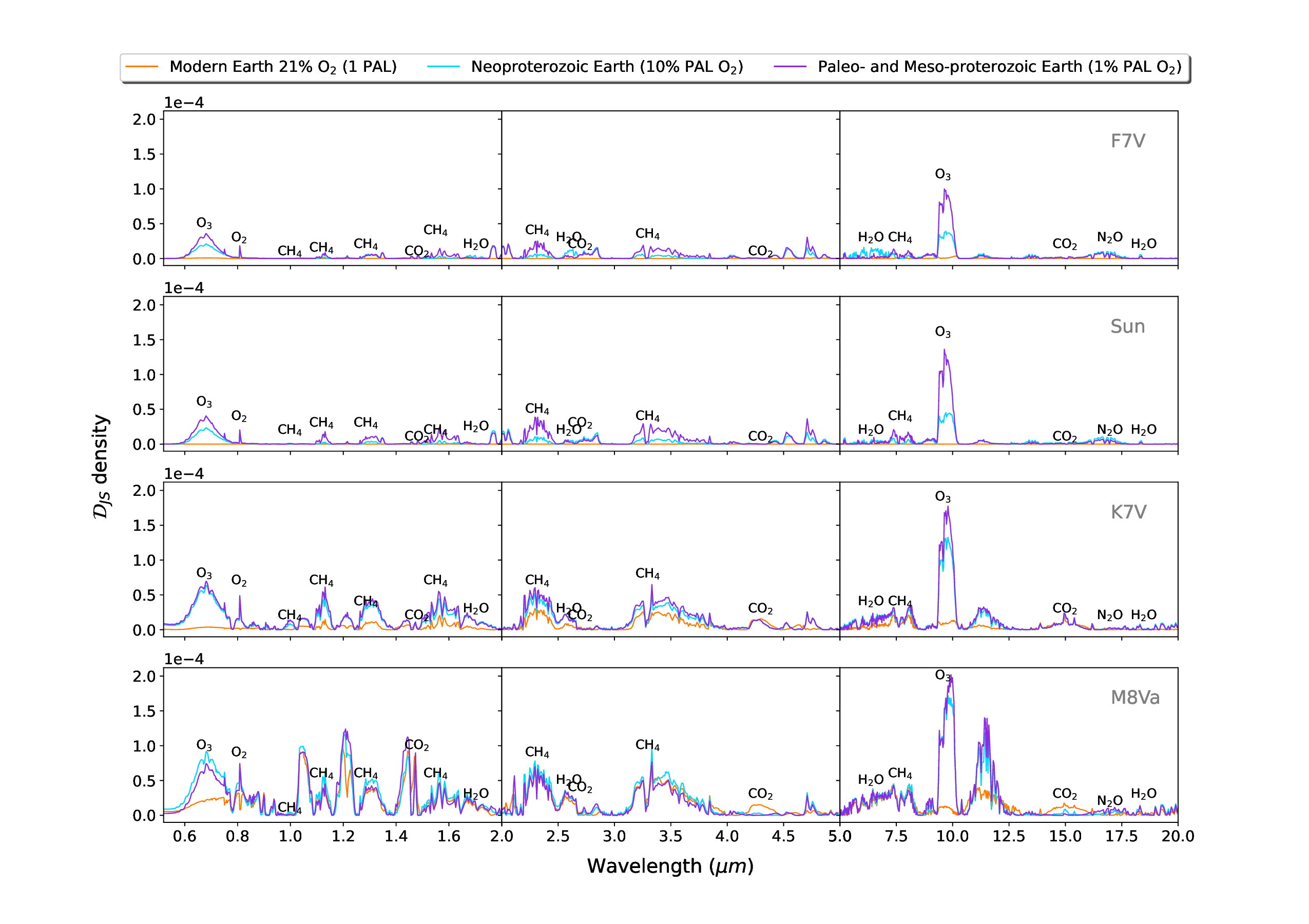}
    \caption{ Jensen-Shannon divergence density ($\djsdens$) of Earth-like planets orbiting different host stars compared to modern Earth. Each row represents Earth-like planets orbiting a different spectral class of host star; each column shows a different wavelength band with higher magnification in the near-infrared to better show the absorption features in that range. We represent $\djsdens$ here as a function of wavelength rather than wavenumber for readability, but emphasize that the two quantities have a one-to-one relationship, so this does not impact our results.}
    \label{fig:djs_dens}
\end{figure*}

\subsection{Identifying Biosignatures}\label{subsec:res2}
As discussed earlier, $\djs$ is a global metric. This means it cannot be used to analyze the information content of individual absorption features. $\djsdens$ provides a valuable companion tool for analyzing the discrete, local spectral features created by potential biosignatures. Figure \ref{fig:djs_dens} shows how $\djsdens$ can be used to identify these spectral features (see e.g. \citep{kaltenegger2017characterize} for more details on the discussion of biosignatures in the atmosphere). For illustrative purposes, we compute the $\djsdens$ of all the systems relative to a modern Earth model around the Sun, a G2V star model. This demonstrates the ability of $\djsdens$ to trace changes in molecular features of the transmission spectrum. However, this analysis shows that the transmission spectrum of an exoplanet should be compared to the modern Earth model of the appropriate spectral class of host star, selected from the bank of simulated transmission spectra like those available from \citet{kaltenegger2020finding} in order to provide an effective tool to identify Earth-analog qualities for real observations. As we seek signs of Earth-like biotic activity, this will allow observers to quantitatively compare an exoplanet's atmospheric spectral features, including potential biosignatures, to that of modern Earth models and to identify similarities. As an example, we analyze two biosignature pairs here: methane (CH$_4$) with ozone (O$_3$); and methane (CH$_4$) with molecular oxygen (O$_2$). 

The changes in the abundance of methane, oxygen, and ozone are clearly visible for different stages of Earth's history in Figure \ref{fig:djs_dens}. Concentrating specifically on the biosoignature pairs (i) CH$_4$ with O$_3$ as well as (ii) CH$_4$ with O$_2$, we discuss the changes in these spectral features below in detail.

Methane: Large methane concentrations for young Earth models result in a large $\djsdens$ relative to modern Earth at methane absorption bands because modern Earth only features a moderate methane content. As the methane mixing ratio reduces, so too does the $\djsdens$ as the transmission spectrum becomes more similar to a modern Earth model. However, the effect of the host star also has a large influence on the analysis: the $\djsdens$ for modern Earth models orbiting host stars with more distant temperatures from the Sun (e.g. K7V and M8Va stars) have lower $\djsdens$ than models for similar epochs for host stars with temperatures more similar to the Sun's (e.g. G2V star). This is shown, for example, in the two O$_3$ features near 0.6$\mu$m and 9.6$\mu$m. The modern Earth model (orange) around the coolest star (M8Va, bottom) has the highest $\djsdens$ of the four stars, followed by the second coolest (K7V, second from bottom). The F7V star (top row) is closest in temperature and UV levels to the Sun (a G2V star) in our grid, giving it a near zero $\djsdens$. Finally, the $\djsdens$ of the modern Earth model around the Sun is zero since the model and comparison planets are identical. 

Oxygen and Ozone: O$_3$ and O$_2$ abundances are highest in modern Earth, progressively decreasing for younger Earth models. As a result, the $\djsdens$ of O$_2$ and O$_3$ are highest for the youngest Earth model, decreasing as the the age of the Earth model approaches 0.0Gya. However, as previously shown, the host star influences the O$_3$ concentration as well as the strength of their spectral features considerably. This means that models for planets orbiting cooler host stars feature higher $\djsdens$ in O$_2$ and O$_3$, especially for younger Earth models (as shown in Figure \ref{fig:djs_dens}). The increase in $\djsdens$ at O$_3$ and O$_2$ absorption bands indicates a net change relative to modern Earth, which could be from an increase \textit{or} a decrease of the absorption feature strength. 

\begin{figure}
    \centering
    \includegraphics[width=\columnwidth]{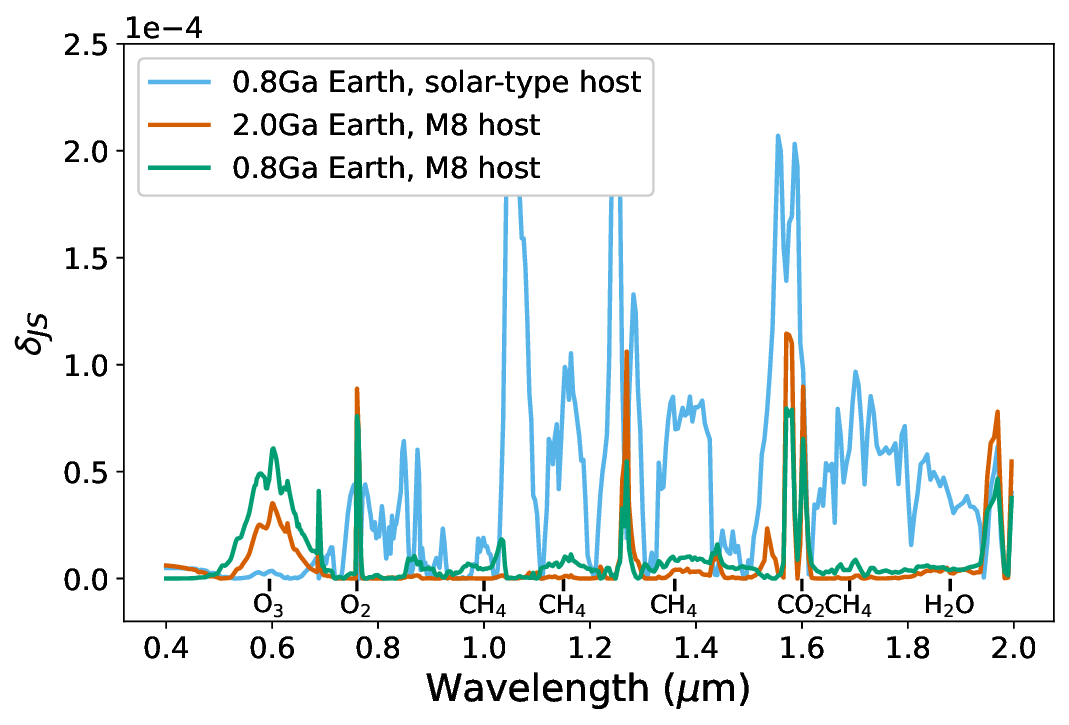}
    \caption{Jensen-Shannon divergence density ($\djsdens$) of the visible range (0.4-2.0$\mu$m) of a modern (0.0Ga) Earth model around an M8 star relative to three different comparison Earth models. Note that the $\djsdens$ for most features is much lower for the Earth models for the correct spectral class of host star (green and orange) than for the solar-type host star (blue), illustrating that this method is useful as a tool for observations when the exoplanet is compared to a model around the appropriate spectral class of host star.}
    \label{fig:M8_comparisons}
\end{figure}

Figure \ref{fig:M8_comparisons} shows how this method can be used as a tool to analyze observations of exoplanets by selecting the comparison Earth models for the correct spectral class host star from the data bank of transmission spectra. The figure shows how the $\djsdens$ decreases for most molecular spectral features when the model transmission spectrum is compared to modern Earth models orbiting similar spectral class stars (green and orange) rather than comparing to the modern Earth analog orbiting the Sun (blue). Once the host star type has been accounted for, $\djsdens$ further decrease (from orange to green) for most features when the age of the comparison planet approaches that of the model planet (from 2.0Ga to 0.8Ga). 

Note that $\djs$ and $\djsdens$ are comparative, rather than \textit{absolute} measures. Because they depend on the resolution, signal-to-noise ratio, and wavelength range of the data, these measures can only be used to rank how ``Earth-analog'' a planet is, rather than act as a binary filter to determine whether a given planet is or is not Earth-analog. In future work, we plan to derive ranges of values of $\djs$ and $\djsdens$ for particular noise and resolution schema of transmission spectra. 

Finally, overlapping features can make it difficult to disentangle which molecule is producing which feature. Since $\djsdens$ uses probability distributions that depend on the full spectrum, the $\djsdens$ depends not just on the spectra at that particular value of $\nu$, but also on the adjacent points in the spectrum. As a result, the metric is sensitive to the shape of the absorption features. This means it could be used to elucidate minor absorption features that may otherwise be obscured.

\section{Discussion}\label{sec:Discussion}
The original analysis of these models in \citet{kaltenegger2020finding} focuses on four different evolutionary epochs---prebiotic Earth 3.9 billion years ago; Earth just after the Great Oxygenation Event 2.0 billion years ago; Earth just after the Neoproterozoic Oxygenation event 0.8 billion years ago; and modern Earth---and provides model spectra for planets orbiting nine different spectral classes of host star from F0 to M8. This database captures the impacts of both the evolution of life and varying spectral irradiance on the spectrum of Earth-like planets. Here, we focus only on the three biotic epochs, as the spectrum of prebiotic Earth is dominated by the methane features rather than by one of the two biosignature pairs we use in our analysis (CH$_4$+O$_2$ and CH$_4$+O$_3$). We plan to add the analysis of atmospheres like Early Mars and prebiotic Earth in our framework in a next paper. We also exclude planets orbiting F0 stars, as the lifespan of these stars is arguably too short for modern Earth life to evolve. In future work, we plan to apply our diagnostic tool to prebiotic Earth models, as well as other lifeless planets test our method with an extended sample of comparison planets. 

The potential biosignatures we have chosen to analyze could be degenerate with imprints from abiotic activity (for more details on our choice, see \citet{kaltenegger2017characterize}). Methane, a key byproduct of anaerobic life, is also produced for example by water-rock reactions on Earth \citep{Schwieterman2018}. Molecular oxygen (O$_2$) could also be produced in higher abundance in a variety of environments, abiotically, especially in planets around cooler stars \citep{clampin2010pathways, tian2014high, gao2015stabilization, luger2015extreme}. However, for now, these biosignature pairs arguably provide the best base for our analysis. Note that the  analysis shown in this paper can easily be extended to other absorption features.

Note that clouds may obscure spectral features if light is unable to penetrate the clouds. On modern Earth, clouds appear primarily below 12km (see e.g. \citet{Betremieux_2014}), which is lower than the effective height of spectral features for a transiting modern Earth \citep{kaltenegger2009detecting}. As a result, clouds on Earth do not obscure spectral features. In addition, clouds only cover about 50\% of modern Earth's surface. However, for cooler host stars light can penetrate deeper into the atmosphere. While there is no consensus on the expected cloud heights of Earth-like planets orbiting different host stars, we follow \citet{kaltenegger2007} and \citet{kaltenegger2020finding} and do not consider cloud obscuration of spectral features in our initial analysis for transiting planets. We plan to expand our analysis for different cloud heights in a future paper.

The sensitivity of our method also depends on the noise and resolution of the data. In \citet{vannah2022informational}, we showed how the sensitivity of the $\djs$ of a transmission spectrum is impacted by Gaussian noise. In future work, we plan to use the JWST simulator JexoSim-2.0 \citep{sarkar_jexosim_2021} to determine the efficacy of our method with realistic noise from JWST. In this analysis, we have reduced the resolution (R=$\lambda $/$\Delta \lambda$) of data to 300, within the range of JWST's NIRSpec and NIRISS spectrographs \citep{2022PASP..134b5002W, 2022A&A...661A..83B, 2022A&A...661A..80J}. 

While we have focused on transmission spectra of Earth-like planets, the framework presented here provides a tool for observers to assess similarities and differences to any chosen planet, in transmission or directly imaged, in emission or reflected light. This approach is not limited to Earth-like planets or to identifying signs of life. This framework can similarly be used to compare any exoplanet spectrum to a Solar System object \citep{Madden2018} or any specifically selected exoplanet to look for similarities.

\section{Conclusions}\label{sec:Conclusions}
We assess whether information theory could identify similarity to modern Earth spectral features in transmission including potential biosignature pairs on Earth-like planets using a model grid of transmission spectra of Earth through time and around different host stars. We defined how ``Earth-analog'' an exoplanet is by the similarity in information content between the transmission spectrum of the model exoplanet and that of modern Earth. We considered three different benchmark epochs in Earth's evolution: a Paleo- to Mesoproterozoic epoch 2.0 billion years ago with 1$\%$ PAL; a Neoproterozoic epoch 0.8 billion years ago with 10$\%$ PAL; and modern Earth. This tool compares the whole spectrum as well as specific wavelength ranges of biosignatures to detect potential signs of life in different context. 

Our analysis has shown that information theory can isolate the information divergence of specific atmospheric spectral features. In this paper we focused, as an example, specifically on two biosignature pairs. Both identifiers used, $\djs$ and $\djsdens$, are sensitive to the shape (amplitude and width) of absorption features, rather than just their presence. Both allow a user to quantitatively assess the likelihood that an exoplanet is Earth-analog with minimal requirements for defining what spectral features to look for.

This approach provides a framework and a tool for observers to assess exoplanet spectra. Just as we analysed simulated transmission spectra of Earth-like planets at different ages and around different host stars in this paper, observed exoplanet spectra could be compared to modern Earth. Additionally, this approach is not limited to Earth-like planets or to identifying signs of life. This framework can similarly be used to compare an exoplanet spectrum to any Solar System object, or any specifically selected exoplanet to look for similarities.

%%%%%%%%%%%%%%%%%%%%%%%%%%%%%%%%%%%%%%%%%%%%%%%%%%
\section*{Data Availability}

The code used to generate the figures and the analysis is available at \href{https://github.com/saracha413/space-djs}{https://github.com/saracha413/space-djs}.
The tools introduced here and in \citet{vannah2022informational} can be used together. $\djs$ can help identify interesting Earth-like planets, creating a pool of candidates. $\djsdens$ can then be used to search through the pool to identify which planets show similar biosignatures to Earth's. 


\bibliographystyle{mnras}
\bibliography{references} % if your bibtex file is called example.bib

\begin{thebibliography}{}
\makeatletter
\relax
\def\mn@urlcharsother{\let\do\@makeother \do\$\do\&\do\#\do\^\do\_\do\%\do\~}
\def\mn@doi{\begingroup\mn@urlcharsother \@ifnextchar [ {\mn@doi@} {\mn@doi@[]}}
\def\mn@doi@[#1]#2{\def\@tempa{#1}\ifx\@tempa\@empty \href {http://dx.doi.org/#2} {doi:#2}\else \href {http://dx.doi.org/#2} {#1}\fi \endgroup}
\def\mn@eprint#1#2{\mn@eprint@#1:#2::\@nil}
\def\mn@eprint@arXiv#1{\href {http://arxiv.org/abs/#1} {{\tt arXiv:#1}}}
\def\mn@eprint@dblp#1{\href {http://dblp.uni-trier.de/rec/bibtex/#1.xml} {dblp:#1}}
\def\mn@eprint@#1:#2:#3:#4\@nil{\def\@tempa {#1}\def\@tempb {#2}\def\@tempc {#3}\ifx \@tempc \@empty \let \@tempc \@tempb \let \@tempb \@tempa \fi \ifx \@tempb \@empty \def\@tempb {arXiv}\fi \@ifundefined {mn@eprint@\@tempb}{\@tempb:\@tempc}{\expandafter \expandafter \csname mn@eprint@\@tempb\endcsname \expandafter{\@tempc}}}

\bibitem[\protect\citeauthoryear{Ahrer et~al.,}{Ahrer et~al.}{2022}]{aher2022}
Ahrer E.-M.,  et~al., 2022, \mn@doi [Nature] {10.1038/s41586-022-05269-w}

\bibitem[\protect\citeauthoryear{Ahrer et~al.,}{Ahrer et~al.}{2023}]{ahrer2023early}
Ahrer E.-M.,  et~al., 2023, \mn@doi [Nature] {https://doi.org/10.1038/s41586-022-05590-4}, pp~1--4

\bibitem[\protect\citeauthoryear{Alderson et~al.,}{Alderson et~al.}{2023}]{alderson2023early}
Alderson L.,  et~al., 2023, \mn@doi [Nature] {https://doi.org/10.1038/s41586-022-05591-3}, 614, 664

\bibitem[\protect\citeauthoryear{Bartlett et~al.,}{Bartlett et~al.}{2022}]{Bartlett2022}
Bartlett S.,  et~al., 2022, \mn@doi [Nature Astronomy] {10.1038/s41550-021-01559-x}, 6, 387

\bibitem[\protect\citeauthoryear{{Birkmann} et~al.,}{{Birkmann} et~al.}{2022}]{2022A&A...661A..83B}
{Birkmann} S.~M.,  et~al., 2022, \mn@doi [\aap] {10.1051/0004-6361/202142592}, \href {https://ui.adsabs.harvard.edu/abs/2022A&A...661A..83B} {661, A83}

\bibitem[\protect\citeauthoryear{Bétrémieux \& Kaltenegger}{Bétrémieux \& Kaltenegger}{2014}]{Betremieux_2014}
Bétrémieux Y.,  Kaltenegger L.,  2014, \mn@doi [The Astrophysical Journal] {10.1088/0004-637X/791/1/7}, 791, 7

\bibitem[\protect\citeauthoryear{Chou et~al.,}{Chou et~al.}{2021}]{Chou2021}
Chou L.,  et~al., 2021, \mn@doi [Frontiers in Astronomy and Space Sciences] {10.3389/fspas.2021.755100}, 8

\bibitem[\protect\citeauthoryear{Clampin}{Clampin}{2010}]{clampin2010pathways}
Clampin M.,  2010, in Pathways Towards Habitable Planets. p.~167

\bibitem[\protect\citeauthoryear{Feinstein et~al.,}{Feinstein et~al.}{2023}]{feinstein2023early}
Feinstein A.~D.,  et~al., 2023, \mn@doi [Nature] {https://doi.org/10.1038/s41586-022-05674-1}, 614, 670

\bibitem[\protect\citeauthoryear{Gao, Hu, Robinson, Li  \& Yung}{Gao et~al.}{2015}]{gao2015stabilization}
Gao P.,  Hu R.,  Robinson T.~D.,  Li C.,   Yung Y.~L.,  2015, Astrophys J, 806, 249

\bibitem[\protect\citeauthoryear{Gardner et~al.,}{Gardner et~al.}{2006}]{gardner2006james}
Gardner J.~P.,  et~al., 2006, \mn@doi [Space Science Reviews] {https://doi.org/10.1007/s11214-006-8315-7}, 123, 485

\bibitem[\protect\citeauthoryear{{Ge} et~al.,}{{Ge} et~al.}{2022}]{ye2022china}
{Ge} J.,  et~al., 2022, {ET White Paper: To Find the First Earth 2.0} (\mn@eprint {arXiv} {2206.06693}), \mn@doi{10.48550/arXiv.2206.06693}

\bibitem[\protect\citeauthoryear{Gleiser \& Stamatopoulos}{Gleiser \& Stamatopoulos}{2012}]{Gleiser2012}
Gleiser M.,  Stamatopoulos N.,  2012, \mn@doi [Physics Letters B] {https://doi.org/10.1016/j.physletb.2012.05.064}, 713, 304

\bibitem[\protect\citeauthoryear{Guttenberg, Chen, Mochizuki  \& Cleaves}{Guttenberg et~al.}{2021}]{Guttenberg2021}
Guttenberg N.,  Chen H.,  Mochizuki T.,   Cleaves H.,  2021, \mn@doi [Life] {10.3390/life11030234}, 11, 234

\bibitem[\protect\citeauthoryear{{Jakobsen} et~al.,}{{Jakobsen} et~al.}{2022}]{2022A&A...661A..80J}
{Jakobsen} P.,  et~al., 2022, \mn@doi [\aap] {10.1051/0004-6361/202142663}, \href {https://ui.adsabs.harvard.edu/abs/2022A&A...661A..80J} {661, A80}

\bibitem[\protect\citeauthoryear{Kaltenegger}{Kaltenegger}{2017}]{kaltenegger2017characterize}
Kaltenegger L.,  2017, \mn@doi [Annual Review of Astronomy and Astrophysics] {https://doi.org/10.1146/annurev-astro-082214-122238}, 55, 433

\bibitem[\protect\citeauthoryear{Kaltenegger \& Sasselov}{Kaltenegger \& Sasselov}{2009}]{kaltenegger2009detecting}
Kaltenegger L.,  Sasselov D.,  2009, \mn@doi [The Astrophysical Journal] {10.1088/0004-637x/708/2/1162}, 708, 1162

\bibitem[\protect\citeauthoryear{{Kaltenegger}, {Traub}  \& {Jucks}}{{Kaltenegger} et~al.}{2007}]{kaltenegger2007}
{Kaltenegger} L.,  {Traub} W.~A.,   {Jucks} K.~W.,  2007, \mn@doi [\apj] {10.1086/510996}, \href {https://ui.adsabs.harvard.edu/abs/2007ApJ...658..598K} {658, 598}

\bibitem[\protect\citeauthoryear{Kaltenegger, Lin  \& Rugheimer}{Kaltenegger et~al.}{2020}]{kaltenegger2020finding}
Kaltenegger L.,  Lin Z.,   Rugheimer S.,  2020, \mn@doi [The Astrophysical Journal] {10.3847/1538-4357/abb9b2}, 904, 10

\bibitem[\protect\citeauthoryear{Kullback \& Leibler}{Kullback \& Leibler}{1951}]{kullback_information_1951}
Kullback S.,  Leibler R.~A.,  1951, \mn@doi [The Annals of Mathematical Statistics] {10.1214/aoms/1177729694}, 22, 79

\bibitem[\protect\citeauthoryear{Lederberg}{Lederberg}{1965}]{Lederberg1965}
Lederberg J.,  1965, \mn@doi [Nature] {https://doi.org/10.1038/207009a0}, 207, 9

\bibitem[\protect\citeauthoryear{Lin}{Lin}{1991}]{lin1991divergence}
Lin J.,  1991, \mn@doi [IEEE Transactions on Information theory] {https://doi.org/10.1109/18.61115}, 37, 145

\bibitem[\protect\citeauthoryear{Lovelock}{Lovelock}{1965}]{Lovelock1965}
Lovelock J.~E.,  1965, \mn@doi [Nature] {https://doi.org/10.1038/207568a0}, 207, 568

\bibitem[\protect\citeauthoryear{Luger \& Barnes}{Luger \& Barnes}{2015}]{luger2015extreme}
Luger R.,  Barnes R.,  2015, \mn@doi [Astrobiology] {10.1089/ast.2014.1231}, 15, 119

\bibitem[\protect\citeauthoryear{Lustig-Yaeger, Robinson  \& Arney}{Lustig-Yaeger et~al.}{2019}]{Lustig2019coronagraph}
Lustig-Yaeger J.,  Robinson T.~D.,   Arney G.,  2019, \mn@doi [Journal of Open Source Software] {10.21105/joss.01387}, 4, 1387

\bibitem[\protect\citeauthoryear{Madden \& Kaltenegger}{Madden \& Kaltenegger}{2018}]{Madden2018}
Madden J.~H.,  Kaltenegger L.,  2018, \mn@doi [Astrobiology] {10.1089/ast.2017.1763}, 18

\bibitem[\protect\citeauthoryear{Madhusudhan, Sarkar, Constantinou, Holmberg, Piette  \& Moses}{Madhusudhan et~al.}{2023}]{madhusudhan2023carbonbearing}
Madhusudhan N.,  Sarkar S.,  Constantinou S.,  Holmberg M.,  Piette A.,   Moses J.~I.,  2023, Carbon-bearing Molecules in a Possible Hycean Atmosphere (\mn@eprint {arXiv} {2309.05566})

\bibitem[\protect\citeauthoryear{Marshall et~al.,}{Marshall et~al.}{2021}]{Marshall2021}
Marshall S.~M.,  et~al., 2021, \mn@doi [Nature Communications] {10.1038/s41467-021-23258-x}, 12, 3033

\bibitem[\protect\citeauthoryear{Pontoppidan et~al.,}{Pontoppidan et~al.}{2022}]{pontoppidan2022jwst}
Pontoppidan K.~M.,  et~al., 2022, \mn@doi [The Astrophysical Journal Letters] {10.3847/2041-8213/ac8a4e}, 936, L14

\bibitem[\protect\citeauthoryear{Ramsay et~al.,}{Ramsay et~al.}{2020}]{ramsay2020eso}
Ramsay S.,  et~al., 2020, in Advances in Optical Astronomical Instrumentation 2019. p. 1120303, \mn@doi{10.1117/12.2541400}

\bibitem[\protect\citeauthoryear{{Robinson}, {Stapelfeldt}  \& {Marley}}{{Robinson} et~al.}{2016}]{Robinson2016}
{Robinson} T.~D.,  {Stapelfeldt} K.~R.,   {Marley} M.~S.,  2016, \mn@doi [\pasp] {10.1088/1538-3873/128/960/025003}, \href {https://ui.adsabs.harvard.edu/abs/2016PASP..128b5003R} {128, 025003}

\bibitem[\protect\citeauthoryear{Rugheimer \& Kaltenegger}{Rugheimer \& Kaltenegger}{2018}]{rugheimer2018spectra}
Rugheimer S.,  Kaltenegger L.,  2018, \mn@doi [The Astrophysical Journal] {10.3847/1538-4357/aaa47a}, 854, 19

\bibitem[\protect\citeauthoryear{Rustamkulov et~al.,}{Rustamkulov et~al.}{2023}]{rustamkulov2023early}
Rustamkulov Z.,  et~al., 2023, \mn@doi [Nature] {https://doi.org/10.1038/s41586-022-05677-y}, pp~1--3

\bibitem[\protect\citeauthoryear{Sarkar \& Madhusudhan}{Sarkar \& Madhusudhan}{2021}]{sarkar_jexosim_2021}
Sarkar S.,  Madhusudhan N.,  2021, \mn@doi [Monthly Notices of the Royal Astronomical Society] {10.1093/mnras/stab2472}, 508, 433

\bibitem[\protect\citeauthoryear{Schwieterman et~al.,}{Schwieterman et~al.}{2018}]{Schwieterman2018}
Schwieterman E.~W.,  et~al., 2018, \mn@doi [Astrobiology] {https://doi.org/10.1089/ast.2017.1729}, 18, 663

\bibitem[\protect\citeauthoryear{Seager}{Seager}{2013}]{seager2013exoplanet}
Seager S.,  2013, \mn@doi [Science] {https://doi.org/10.1126/science.1232226}, 340, 577

\bibitem[\protect\citeauthoryear{Seager}{Seager}{2014}]{seager2014future}
Seager S.,  2014, \mn@doi [Proceedings of the National Academy of Sciences] {https://doi.org/10.1073/pnas.1304213111}, 111, 12634

\bibitem[\protect\citeauthoryear{Shannon}{Shannon}{1948}]{Shannon1948}
Shannon C.~E.,  1948, \mn@doi [Bell System Technical Journal] {10.1002/j.1538-7305.1948.tb00917.x}, 27, 623

\bibitem[\protect\citeauthoryear{Tian, France, Linsky, Mauas  \& Vieytes}{Tian et~al.}{2014}]{tian2014high}
Tian F.,  France K.,  Linsky J.~L.,  Mauas P.~J.,   Vieytes M.~C.,  2014, \mn@doi [Earth and Planetary Science Letters] {10.1016/j.epsl.2013.10.024}, 385, 22

\bibitem[\protect\citeauthoryear{Tinetti et~al.,}{Tinetti et~al.}{2018}]{tinetti2018chemical}
Tinetti G.,  et~al., 2018, \mn@doi [Experimental Astronomy] {https://doi.org/10.1007/s10686-018-9598-x}, 46, 135

\bibitem[\protect\citeauthoryear{{Traub} \& {Jucks}}{{Traub} \& {Jucks}}{2002}]{traub2002}
{Traub} W.~A.,  {Jucks} K.~W.,  2002, \mn@doi [Geophysical Monograph Series] {10.1029/130GM25}, \href {https://ui.adsabs.harvard.edu/abs/2002GMS...130..369T} {130, 369}

\bibitem[\protect\citeauthoryear{Vannah, Stiehl  \& Gleiser}{Vannah et~al.}{2022}]{vannah2022informational}
Vannah S.,  Stiehl I.~D.,   Gleiser M.,  2022, An Informational Approach to Exoplanet Characterization (\mn@eprint {arXiv} {2206.13344})

\bibitem[\protect\citeauthoryear{{Willott} et~al.,}{{Willott} et~al.}{2022}]{2022PASP..134b5002W}
{Willott} C.~J.,  et~al., 2022, \mn@doi [\pasp] {10.1088/1538-3873/ac5158}, \href {https://ui.adsabs.harvard.edu/abs/2022PASP..134b5002W} {134, 025002}

\bibitem[\protect\citeauthoryear{Wong, Prabhu, Williams, Morrison  \& Hazen}{Wong et~al.}{2023}]{Wong2023}
Wong M.~L.,  Prabhu A.,  Williams J.,  Morrison S.~M.,   Hazen R.~M.,  2023, \mn@doi [Journal of Geophysical Research: Planets] {10.1029/2022JE007658}, 128

\makeatother
\end{thebibliography}

%\begin{thebibliography}{99}
%\end{thebibliography}

% Don't change these lines
\bsp	% typesetting comment
\label{lastpage}
\end{document}

%\maketitle
\begin{figure*}
    \centering
    \includegraphics[width=\textwidth]{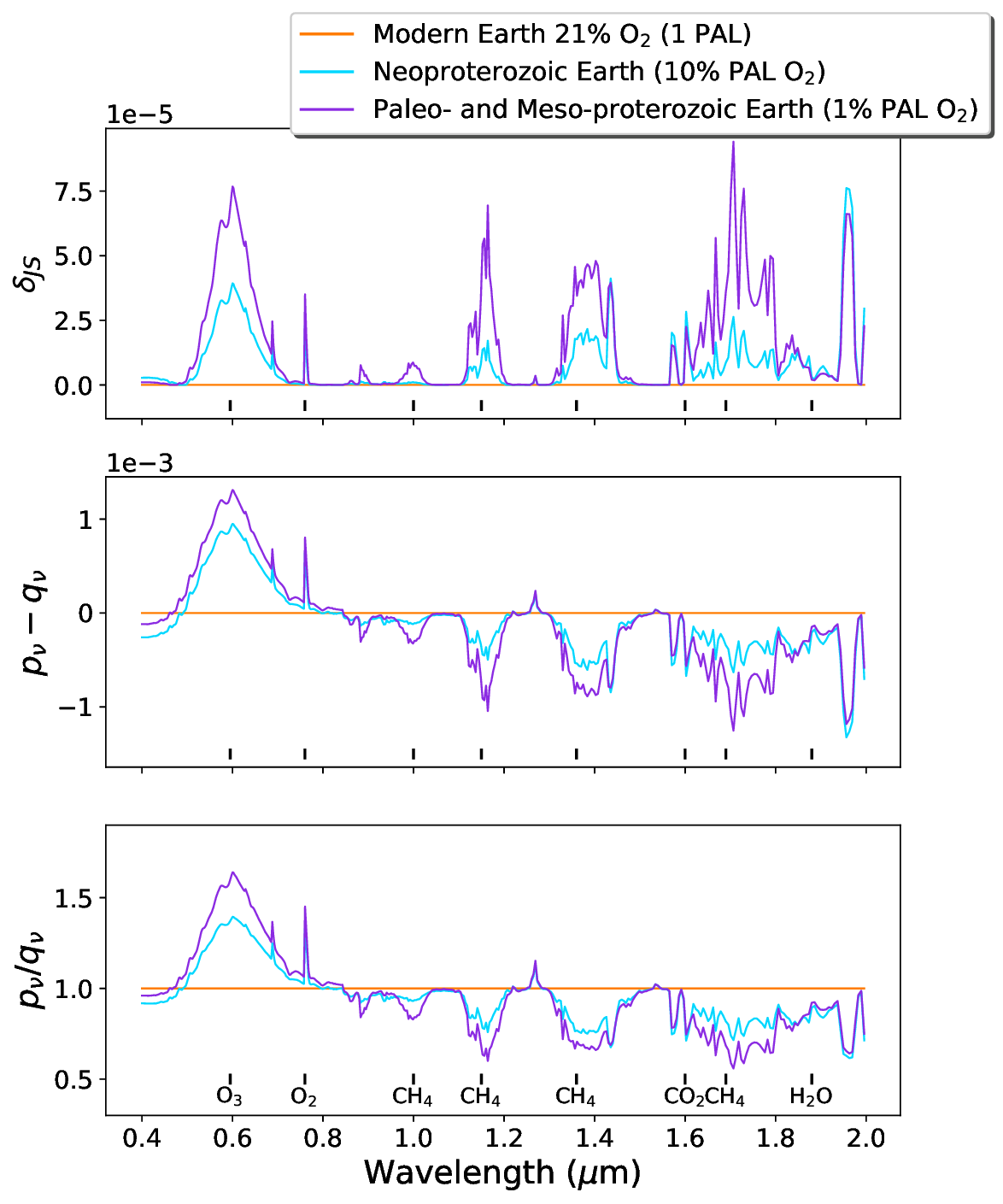}
    \caption{For comparison, we include plots showing how $\djsdens$ (top row) compares to the simpler functions $p_\nu-q_\nu$ (middle row) and $p_\nu/q_\nu$ (bottom row). We compare only the visible range (0.4-2.0$\mu m$), and compare each planet to a modern Earth model around our Sun, as in the second row of Figure 2. We observe first that the two simple functions are not metrics: $p_\nu-q_\nu$ may be less than zero, while $p_\nu/q_\nu$ is not equal to 0 when $p_\nu=q_\nu$ (orange line). Simple modifications to these functions such as $|p_\nu-q_\nu|$ also fail to produce true metrics (for example, $|p_\nu-q_\nu|$ does not obey the triangle inequality). This is key: as a metric, $\djsdens$ represents a distance in information space, while the non-metric functions only represent a difference between the modal fractions at particular values of $\nu$. As expected, the simple functions and $\djsdens$ generally show similar results. $\djsdens$ appears less dependant on the spectral tilt on the blue end of the visible range, but more sensitive to absorption features such as the twin peaks of the CO$_2$ feature at 1.6$\mu$m. }
    \label{fig:M8_comparisons}
\end{figure*}

\begin{figure*}
    \centering
    \includegraphics[width=\textwidth]{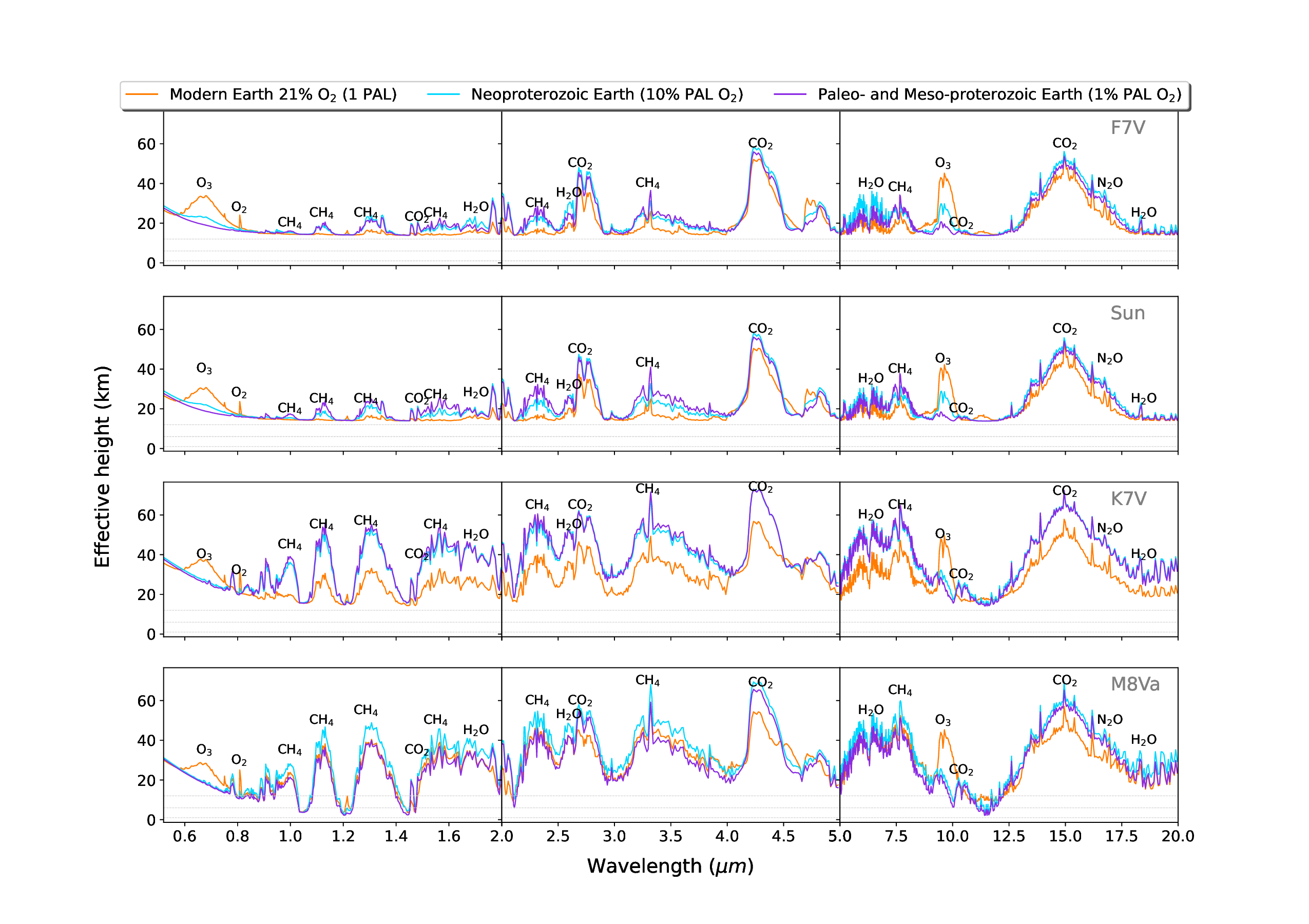}
    \caption{Raw data from Kaltenegger et al. (2020) used to produce the results for this work, smoothed to a resolving power (R=$\lambda $/$\Delta \lambda$) of 300 using a top-hat convolution. Each row shows the transmission spectrum of an Earth model orbiting a different spectral class of host star, labeled in gray. Each line represents the transmission spectrum of the Earth clone at a different age. The thin, gray dashed lines represent potential cloud heights of 6, 9, and 12km.}
    \label{fig:djs_time}
\end{figure*}